\documentclass[12pt,reqno]{amsart}

\usepackage[margin=2.5cm]{geometry}
\usepackage{tikz}
\tikzset{node distance=2cm, auto}
\usetikzlibrary{matrix,arrows,decorations.pathmorphing,decorations.markings,calc}

\newcommand\arXiv[1]{\href{http://arxiv.org/abs/#1}{\nolinkurl{arXiv:#1}}}

\usepackage{caption}
\usepackage{graphicx}
\usepackage{subfig}
\usepackage{rotating}

\usepackage{listings}
\usetikzlibrary{angles}
\usetikzlibrary{quotes}
\usepackage{float}
\usepackage{amsmath,amssymb}

\usepackage{amscd}
\usepackage{verbatim}
\usepackage{enumerate}
\usepackage{bbm}
\usepackage{mathrsfs}
\usepackage{color}
\newtheorem{theorem}{Theorem}

\newtheorem{problem}{Problem}

\newcommand{\lob}{\Lambda}

\newcommand{\be}{\begin{equation}}
\newcommand{\ee}{\end{equation}}

\newcommand{\x}{\times}

\title{Abelian Higgs Vortices and Discrete Conformal Maps}

\author{Alexander I. Bobenko}
\address{A.B.: Institut f{\"u}r Mathematik, Technische Universit{\"a}t Berlin, Strasse des 17. June 136, 10623 Berlin,
Germany.}
\email{E-mail: bobenko@math.tu-berlin.de}

\author{Ananth Sridhar}
\address{A.S.: Institut f{\"u}r Mathematik, Technische Universit{\"a}t Berlin, Strasse des 17. June 136, 10623 Berlin,
Germany.}
\email{E-mail: sridhar@math.tu-berlin.de}

\begin{document}

\begin{abstract} 
We establish a connection between recent developments in the study of vortices in the abelian Higgs models, and in the theory of structure-preserving discrete conformal maps. We explain how both are related via conformal mapping problems involving prescribed linear combinations of the curvature and volume form, and show how the discrete conformal theory can be used to construct discrete vortex solutions.
\end{abstract}

\maketitle

\section*{Introduction}
An important class of problems in mathematics take the following prototypical form: given a Riemmanian manifold, find a conformally equivalent metric satisfying certain prescribed properties. Such \emph{conformal mapping problems} arise ubiquitiously, and their solutions are invaluable to many areas of applied mathematics, physics, and engineering.

Inspired largely by these applications, considerable work has been done in developing notions of \emph{discrete conformal maps}. Of particular interest have been discretizations that preserve certain structural properties and characteristics of their continuum counterparts. Such discretizations have shown to contain a surprisingly profound and rich theory of their own, and their study is an exciting and flourishing area of mathematics today.

The main purpose of this paper is to note a correspondence between continuum conformal mapping problems that arise in the study of vortices in the abelian Higgs field theory (see \cite{Ba,M}) and certain discrete conformal mapping problems studied in \cite{BPS,BSS}.

The abelian Higgs field theory first arose as a phenomenological model of superconductivity. The model admits vortex solutions, which are topogically stable, finite energy solutions to the field equations. Physically, the vortices are crucial in understanding the Meissner effect of magnetic repulsion in superconductors. It was observed by Witten \cite{W} that in some circumstances, the equations describing vortices are closely related to the Liouville equation for the conformal factor of constant curvature surfaces, which he exploited to construct explicit vortex solutions. This was further generalized by recent work by \cite{Ba,M}, who showed that the vortex equations can be reformulated as a conformal mapping problem involving prescribed linear combinations of the Gaussian curvature and the Riemannian volume form.

On the other hand, \cite{BPS} studied discrete conformal maps of triangulations, in which metrics correspond to edge lengths and conformal transformations to scale factors associated with vertices. These notions were generalized to hyperbolic and spherical triangulations (see \cite{BPS,BSS}). Several types of associated mapping problems were studied and shown to be solvable by variational principles. We argue that these discrete conformal problems correspond naturally to continuum problems involving combinations of the curvature and volume form. We then explain how the discrete conformal theory can be used to construct discrete vortex solutions.

\begin{figure}[h]
\subfloat[Subfigure 1 list of figures text][]{
\includegraphics[scale=.5]{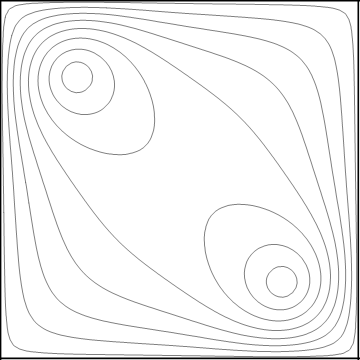}
\label{fig:vortex1}} \;\;\;\;\;\;
\subfloat[Subfigure 2 list of figures text][]{
\includegraphics[scale=.5]{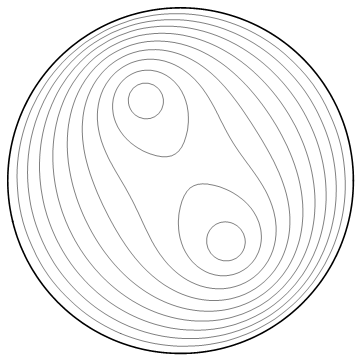}
\label{fig:vortex2} }
\caption{Equidistant level curves of the Higgs field of two Taubes vortices (A) on a square and (B) on a disk. Compare to \cite{N}, Figure 5. } \label{fig:vortices}\end{figure}

Acknowledgements: We graciously thank Thilo R{\"o}rig for help producing Figure \ref{fig:vortices} using the software VaryLab \cite{VL}. This research was supported by the DFG Collaborative
Research Center TRR 109 ``Discretization in Geometry and Dynamics.''

\section{Vortices in the Abelian Higgs Theory}

\subsection{The Abelian Higgs Action}
The classical abelian Higgs field theory is defined on a hermitian line bundle $ L \rightarrow M $ over a closed, oriented two dimensional Riemannian manifold $ M $. The fields are a section $\phi $ of $ L $, and a Hermitian $U(1) $-connection $ A $ on $ L $.

The action functional of the Higgs field theory is given by
\begin{align} \label{eq:action}
S[\phi, A] =  \int_M \left( | F_A|^2 - 2 C \; | d_A \phi |^2 + \left(C | \phi |^2 - C_0 \right)^2 \right) \; dV,
\end{align}
where $ C $ and $ C_0 $ are real constants, $ F_A \in \Omega^2(M) $ is the curvature of the connection, and $ d_A: \Omega^k(M, L) \rightarrow \Omega^{k+1} (M, L) $ is the covariant derivative, and $ dA $ is the Riemannian volume element.

In a trivializing coordinate chart, the connection can be represented by a differential one form $ A \in \Omega^1(M) $. The curvature two form is $ F_A = dA $, and the covariant derivative $ d_A \phi = d\phi - i  A \phi $. The Euler-Lagrange equations are given by
\begin{equation}\label{eq:EL}
\begin{aligned}
\star d \star dA - i C \left(\bar{\phi} \; d_A \phi - \phi \; \overline{d_A \phi}\right) &= 0, \\
 \; \star d_A \star d_A \phi - \; (C \; | \phi|^2 - C_0) \phi &= 0, 
\end{aligned}
\end{equation}
where $ \star $ is the Hodge star.

\subsection{Vortex Equations}
The vortex equations are a set of first order equations describing stationary points of the abelian Higgs action; when $ C < 0 $, solutions of the vortex equations are global minima of the action functional. The equations are found after rearranging the action by exploiting the complex structure on $ M $. 

Recall that the Hodge star induces a complex structure and gives a decomposition of complexified forms $ \Omega^1(M,\mathbb{C}) = \Omega^{1,0}(M,\mathbb{C}) \oplus \Omega^{0,1}(M,\mathbb{C}) $, and of the exterior derivative $ d = \partial + \bar \partial $. There is correspondingly a decomposition $ \Omega^1(M,P) = \Omega^{1,0}(M,P) \oplus \Omega^{0,1}(M,P) $ and a splitting $ d_A = \partial_A + \bar \partial_A $. In addition, denote by $ \Lambda: \Omega^{k} \rightarrow \Omega^{k-2} $ the contraction with volume form; on a two dimensional manifold, $ \Lambda \omega = \langle \omega, \Lambda \rangle $ for any $ \omega \in \Omega^2 $, and $ \langle , \rangle $ is the pointwise inner product on forms induced by the Riemannian metric. 

The action (\ref{eq:action}) can then be rewritten (see \cite{Br,M}) as
\begin{align}
S[\phi, A] &= \int_M \left( \left( \Lambda F_A + (C_0 - C | \phi |^2) \right)^2  - C | \bar \partial_A \phi | ^ 2 \right)\; dV + \int_M C_0 F_A.
\end{align}
Since the integral of the curvature is quantized and independent of the connection
\begin{align*}
\int_M F_A = 2 \pi N,
\end{align*}
where $ N$ is the Chern number and an invariant of the bundle $ P $, the action can then be written
\begin{align}
S[\phi,A ]= \int_M \left( \left( \Lambda F_A + (C_0 - C | \phi |^2) \right)^2  - 2 C | \bar \partial_A \phi | ^ 2 \right) \; dV + 4 \pi N C_0. \label{eq:action2}
\end{align}

The \emph{vortex equations} demand that the each term of the integrand vanishes:
\begin{align}
\bar\partial_A \phi &= 0, \label{eq:vorteq1} \\
\Lambda F_A &= -C_0 + C | \phi |^2. \label{eq:vorteq2}
\end{align}
\noindent The vortex equations imply the Euler-Lagrange (\ref{eq:EL}) equations. Moreover, when $ C \leq 0 $, each term of the integrand  (\ref{eq:action2}) is positive definite, from which it follows that solutions to the vortex equations minimize the action.

The first vortex equation (\ref{eq:vorteq1}) says that $ \phi $ is gauge-covariantly holomorphic. It is convenient to assume complex coordinates $ (z, \bar{z})$, in which the metric takes the conformal form
\begin{align} \label{eq:isomet}
ds^2 = e^{2 \rho}\; dz \; d \bar{z}.
\end{align}
The connection can be written $ A = A_z \; dz + A_{\bar{z}} \; d\bar{z} $, where by unitarity, $ \overline{A_z} = A_{\bar{z}} $. 

In these coordinates, the equation (\ref{eq:vorteq1}) can be expanded as
\begin{align*}
\bar \partial \phi - i \phi A_{\bar{z}} &= 0,
\end{align*}
and be integrated
\begin{align*}
A_{\bar{z}} = - i \bar \partial \log(\phi).
\end{align*}
This yields for the curvature
\begin{align}\label{eq:vorcurv}
F = dA = -i \; \partial \bar \partial \log | \phi | ^2 \; dz \wedge d\bar{z}.
\end{align}

Since $ \phi $ is gauge covariantly holomorphic, it has isolated zeroes at points $ z_i $, called the \emph{vortex centers}, with multiplicities $ n_i $, called \emph{vortex numbers}. From (\ref{eq:vorcurv}), near each vortex center $ z_i $, the curvature has a singularity as $ z \rightarrow z_i $ of the form
\begin{align*}
F \sim  n_i \;  \partial \bar \partial \log( z - z_i )  = n_i \; \delta(z - z_i).
\end{align*}
It follows that the total vortex number is given by
\begin{align} \label{eq:vortnumber}
\int_M F = 2 \pi \sum n_j = 2 \pi N.
\end{align}

Turning to the second vortex equation (\ref{eq:vorteq2}), using $ (\ref{eq:vorcurv})$, one finds
\begin{align*}
\Delta \log( |\phi|^2 ) = C_0 - C | \phi |^2 ,
\end{align*}
where $ \Delta = e^{-2 \rho} \; \partial \bar \partial $ is the Laplace-Beltrami operator. In terms of the function $u $ defined by $ e^{2 u} = | \phi| ^2 $, this can be written as
\begin{align} \label{eq:vortLiouville}
\Delta u  =  C_0 - C e^{2 u} .
\end{align}

\subsection{Five Vortex Equations}
By rescaling first the metric and then shifting $ u $ (rescaling $ \phi $), the constants $ C $ and $ C_0 $ can without loss of generality be assumed to take the values $ \{ -1, 0, +1 \} $, giving nine vortex equations. However, since $ \phi $ is covariantly holomorphic and has only zeroes, (\ref{eq:vortnumber}) implies that $ N > 0 $. Then, since by (\ref{eq:vorteq2})
\begin{align*}
0 < 2 \pi N = \int_M F = \int  \left( -C_0 + C e^{2 u} \right) \; dV,
\end{align*}
it follows that solutions only exist if the integrand on the right side can be positive. This discludes the four cases $ C_0 \in \{0,1\}, C \in \{-1,0\} $, and yields five remaining vortex equations:

\begin{enumerate}
\item $ C_0 = -1, C = -1 $. Taubes Vortex. 
\item $ C_0 =  -1, C = 0 $. Bradlow Vortex. 
\item $ C_0 = -1, C = 1 $.  Ambjørn-Olesen Vortex. (see \cite{AO})
\item $ C_0 = 0, C = 1 $. Jackiw-Pi Vortex.
\item $ C_0 = 1, C= 1 $. Popov Vortex.
\end{enumerate}

\subsection{The Baptista Metric}
The Gaussian curvature $ K $ under conformal scaling of the metric $ ds^2 \rightarrow e^{2 u} ds^2 $ transforms as
\begin{align}
K \rightarrow e^{-2 u} ( K + \Delta u ) .
\end{align}
So the Taubes vortex equation corresponds to the Liouville equation in constant negative curvature.

This was pursued further in \cite{Ba}, who study the conformally rescaled metric
\begin{align*}
ds_B^2 = e^{2 u} ds^2.
\end{align*}
The Baptista metric is degenerate, with cone type singularities at the vortex centers where $ e^{2 u} $ vanishes, with excess angle $ 2 \pi n_i $.

The curvature $ K_B $ of the Baptista metric away from the vortex centers can be computed using Liouville's equation, and using the vortex equation (\ref{eq:vortLiouville} for $ \Delta u $ to obtain
\begin{align*}
e^{2 u} K_B &= K - (C_0 - C e^{2 u}).
\end{align*}
Rearranging gives
\begin{align*}
( K_B - C ) e^{ 2 u} = K - C_0 .
\end{align*}
Denoting by $ dV_B $ and $ dV $ the volume forms, the Baptista equation can equivalently be written as
\begin{align} \label{eq:bapt}
(K_B- C) \; dV_B = (K - C_0) \; dV.
\end{align}
A field configuration $ (A, \phi) $ is a solution to the vortex equations if and only if the associated Baptista metric satisfies (\ref{eq:bapt}) away from the vortex centers. 

\section{Discrete Conformal Maps}

\subsection{Triangulated Surfaces and Discrete Metrics}
A \emph{triangulated surface} consists of a compact oriented surface $ S $, a finite simplicial complex $ \Sigma $, and a homeomorphism $ h: \Sigma \rightarrow S $. Combinatorially, the complex $ \Sigma $ consists of vertices $V$, edges $E$, and faces $ F $. We will enumerate vertices, and denote by $ ij $ the edge adjacent to vertices $ i $ and $ j $, and by $ ijk $ the face adjacent to vertices $ i, j $ and $ k $. 

A \emph{geodesically triangulated surface} is a triangulated surface such that image of each edge of $ \Sigma $ under the homeomorphism $h$ is geodesic in $ S $. A \emph{euclidean triangulated surface} is a geodesically triangulated surface with a smooth Riemannian metric defined on $ S \backslash V $ such each point of $ S \backslash V $ is locally isometric to the Euclidean plane. In a \emph{hyperbolic triangulated surface}, each point of $ S \backslash V $ is locally isometric to the hyperbolic plane.  In a \emph{spherical triangulated surface}, each point of $ S \backslash V $ is locally isometric to the sphere. 

Such triangulated surfaces can be obtained by gluing a finite collection of either euclidean, hyperbolic, or spherical triangles along edges. These surfaces have constant Gaussian curvature except at vertices where, the metric may have a conical singularity. We denote by $ \Theta_i $ the total angle about vertex $ i \in V$.  

A discrete metric on $ \Sigma $ is the pair $ (l, g) $, where $ g \in \{\text{euclidean}, \text{hyperbolic}, \text{spherical} \} $ describes the geometry of the faces, and $ l: E \rightarrow \mathbb{R}_{\geq 0} $ the length of edges. The map $ l $ satisfies the triangle inequalities
\begin{align}\label{eq:tri}
l_{ij} + l_{jk} < l_{ik} \hspace{20pt} \forall \;  ijk \in F.
\end{align}
In addition, for the case of spherical geometry $  l $ must satisfy 
\begin{align} \label{eq:sphc}
l_{ij}+l_{ik} + l_{jk} < 2 \pi \hspace{20pt} \forall \;  ijk \in F.
\end{align}
A discrete metric satisfying these inequalities defines a unique triangulated surface that we will denote $ (\Sigma, l,g) $. 

\subsection{Conformal Equivalence of Discrete Metrics}
\begin{figure}[h]
\subfloat[Subfigure 1 list of figures text][]{
\begin{tikzpicture}
\draw[scale=1,domain=0:360,smooth,variable=\x,black] plot ({cos(\x)},{sin(\x)});
\draw[scale=1,domain=140:40,smooth,variable=\x,black, ultra thick] plot ({cos(\x)},{sin(\x)});
\draw[dashed] (-0.766044,0.642788) -- (0.766044,0.642788);
\node at (0,1.2){$ l$};
\node at (0,.4){$ L$};
\end{tikzpicture}\label{fig:sphere}} \;\;\;\;\;\;
\subfloat[Subfigure 2 list of figures text][]{
\begin{tikzpicture}[scale = .55]
\draw[domain=-4:4,smooth,variable=\x,black] plot ({\x},{sqrt(1 + \x*\x)});
\draw[domain=-2:2,smooth,variable=\x,black, ultra thick] plot ({\x},{sqrt(1 + \x*\x)});
\draw[dashed] (-2, 2.23607) -- (2, 2.23607);
\node at (0,.5){$ l$};
\node at (0,2.6){$ L$};
\end{tikzpicture}\label{fig:hyper}}
\caption{The intrinsic distance $ l $ and extrinsic distance $ L $ for the spherical and hyperbolic cases.}
\label{fig:Ll}
\end{figure}
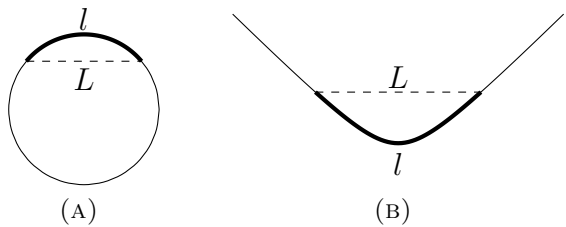

The distance function $ L: E  \rightarrow \mathbb{R}_{\geq 0} $ on a triangulated surface $ (\Sigma, l)_g $ is defined as
\begin{align*}
L_{ij} = \begin{cases}l_{ij} &\text{ if $ g $ is euclidean}, \\
\sinh \frac{l_{ij}}{2} &\text{ if $ g $ is hyperbolic}, \\
\sin \frac{l_{ij}}{2} &\text{ if $ g $ is spherical}.
\end{cases}
\end{align*}
The distance function can be understood by imagining the hyperbolic plane and sphere as isometrically embedded in $ \mathbb{R}^{2,1} $ and $ \mathbb{R}^{3} $ respectively; then $ L$ is the ambient distance of two points seperated by intrinsic distance $ l $, see Figure \ref{fig:Ll}.

Two discrete metrics $ (l,g) $ and $ (\tilde{l},\tilde{g}) $ are \emph{discrete conformally equivalent} if there exists a function (logarthmic scale factor) $ u: V \rightarrow \mathbb{R} $ such that
\begin{align}\label{eq:confrmequiv}
\widetilde{L}_{ij} = e^{\frac{1}{2} \left( u_i + u_j \right) } L_{ij}.
\end{align}\noindent Note that we do not require the geometries $ g $ and $ \tilde{g} $ to be the same. For the euclidean geometry, this definition first appeared in \cite{Luo}, for the spherical and hyperbolic geometry in \cite{BPS}, and in all generality in \cite{BSS}.

\subsection{The Discrete Conformal Mapping Problem and Continuum Counterparts}
In \cite{BSS}, several conformal mapping and uniformization problems were studied in detail. In this note, we focus in particular on the following special case:

\begin{problem}[The Discrete Conformal Mapping Problem]\label{cmp}
Given a triangulated surface with discrete metric $ (l, g) $, find a conformally equivalent metric $ (\widetilde{l}, \widetilde{g} ) $ with prescribed conical angle $ \widetilde \Theta: V \rightarrow \mathbb{R} $ and prescribed geometry $ \widetilde{g} \in $ \{euclidean, hyperbolic, spherical\}. 
\end{problem}

Let us now describe the analagous continuum conformal mapping problem. 
The conical defect angle has often been considered as a discretization of the Guassian curvature. More precisely (see \cite{BPS}), for a Euclidean triangulation, the defect angle is a discretization of curvature two form $ K \; dV $. This is evident from the scaling behaviour: upon rescaling, both the curvature form and angular defect are preserved. For the more general case of hyperbolic and spherical triangulations, the conical defect angle should be understood as a discretization of a linear combination of the curvature two form and the volume form. This is motivated as follows:

Consider the Voronoi cell $ U $ of a vertex $ i \in V $. It consists of a finite number of geodesic triangles $ T_i $ for $ i = 1 \cdots n $, and has a piecewise geodesic boundary $ \partial U $ consisting of smooth segments $ \partial U_i $ as shown in Figure \ref{fig:vor}.

\begin{figure}[h]
\begin{tikzpicture}[scale = 2.8]
\coordinate (p0) at  (-1,.1);
\coordinate (p1) at (-.6,-.5);
\coordinate (p2) at (0,-.8) ;
\coordinate (p3) at (.6,-.3);
\coordinate (p4) at (.5,.5);
\coordinate (p5) at (-.4,.8);
\coordinate (c0) at (0,0);
\coordinate (e0) at (0.3, -0.95);

\draw (p0)--(p1)--(p2)--(p3)--(p4)--(p5)--(p0);
\draw (c0)--(p0);
\draw (c0)--(p1);
\draw (c0)--(p2);
\draw (c0)--(p3);
\draw (c0)--(p4);
\draw (c0)--(p5);
\draw [dashed] (p2)--(e0);

\pic [draw, "\tiny{$\gamma_i$}",angle radius = .55cm] { angle = p1--c0--p2};
\pic [draw, "\tiny{$\alpha_i$}",angle radius = .55cm] { angle = c0--p2--p1};
\pic [draw, "\tiny{$\beta_i$}",angle radius = .55cm] { angle = p2--p1--c0};
\pic [draw, "\tiny{$\theta_i$}",angle radius = .7cm] { angle = e0--p2--p3};
\end{tikzpicture} \caption{} \label{fig:vor}
\end{figure}

The integral of the curvature form over $ U $ is given by the Gauss-Bonnet theorem, including terms accounting for boundary as
\begin{align*}
\int_U K\;  dV  &= 2 \pi \chi(U) - \sum_i \int_{\partial U_i} k_g - \sum_{i} \theta_k \\
&= 2 \pi - \sum_i \theta_i.
\end{align*}
Here $ \theta_i $ is the turning angle between the adjacent segments $  \partial U_{i} $ and $ \partial U_{i+1} $ (see Figure \ref{fig:vor}). Rewriting in terms of internal angles gives:
\begin{align}
\sum_i \theta_i &= \sum \left( \pi - (\alpha_i + \beta_{i+1} ) \right) =\sum_k \left( \pi  - \alpha_i - \beta_i  \right) \label{eq:sum2}
\end{align}
For spherical/hyperbolic triangles, the area $ \Delta $ of a triangle is related to the angles of the triangle by
\begin{align*}
\Delta = \pm ( \alpha + \beta + \gamma - \pi ),
\end{align*}
so that in any case
\begin{align*}
(\pi - \alpha - \beta ) = \gamma - C_g \; \Delta,
\end{align*}
where $ C_g  = +1, 0, -1 $ for $ g = $ spherical, euclidean, and hyperbolic respectively.

Substituting this expression into the sum (\ref{eq:sum2}) gives:
\begin{align*}
\int_U K \; dV &= \sum_i \gamma - C_g \; \Delta = \Theta - \int_U C_g dV,
\end{align*}
or upon rearranging
\begin{align*}
\Theta  = \int_{U} \left( K + C_g \right) \; dV.
\end{align*}

Thus, the conical angle $ \Theta $ corresponds most naturally to the two form $ (K + C_g) \; dV$.  The analogous continuous mapping problem is:
\begin{problem}[The Continuum Conformal Mapping Problem] \label{ccmp}
Given a Riemannian manifold $ (M,g) $, find a conformally equivalent metric with prescribed linear combination of curvature form and area form
\begin{align*}
(K + C) \; dV,
\end{align*}
where $ C \in \{-1,0,1\} $.
\end{problem}

\subsection{The Variational Principle for the Discrete Conformal Mapping Problem}
The discrete conformal mapping problem \ref{cmp} can be solved by variational principles that we briefly review now (see \cite{BSS,BPS} for further details).

Fix a triangulated surface $ ( \Sigma, l, g) $, along with a target geometry $ \widetilde{g} $ and target angles $ \Theta: V \rightarrow \mathbb{R} $. The \emph{feasible region} $ \mathcal{F} $ is the subset of logarthmic conformal factors $ \{ u : V \rightarrow \mathbb{R} \} $ such that conformally transformed $ \widetilde{l}$ defined via equation (\ref{eq:confrmequiv}) defines a discrete metric in the $ \tilde{g} $ geometry; i.e. $ \widetilde{l} $ should satisfy the triangle inequalities (\ref{eq:tri}), and also inequality (\ref{eq:sphc}) in the case that $ \widetilde{g} $ is spherical.

It is convenient to introduce the logarithmic lengths
\begin{align*}
\lambda_{ij} = \log L_{ij}.
\end{align*}
In terms of $ \lambda$, equation (\ref{eq:confrmequiv}) takes the form
\begin{align} \label{eq:lambdaequiv}
\widetilde{\lambda}_{ij} = u_i + u_j + \lambda_{ij}.
\end{align}

The functional $ E^{\tilde{g}}_{\Theta}: \mathcal{F} \rightarrow \mathbb{R} $ is defined as
\begin{align*}
E^{\tilde{g}}_{\Theta}(u) = \sum_{ijk\in F} \left( f^{\widetilde{g}}(\widetilde{\lambda}_{ij}, \widetilde{\lambda}_{jk}, \widetilde{\lambda}_{ki}) - \frac{\pi}{2}(\widetilde{\lambda}_{jk} + \widetilde{\lambda}_{ki} + \widetilde{\lambda}_{ij} ) \right)  + \sum_{i \in V} \Theta_i u_i .
\end{align*}
The function $ f^{\widetilde{g}} $ here is defined as
\begin{align*}
f^{\widetilde{g}}(\lambda_1, \lambda_2, \lambda_3) =& \; \beta_1 \lambda_1 + \beta_2 \lambda_2 + \beta_3 \lambda_3 + \lob(\alpha_1) + \lob(\alpha_2) + \lob(\alpha_3) \\ &+ \lob( \beta_1) + \lob(\beta_2)  + \lob(\beta_3) + \lob\left( \frac{1}{2} ( \pi - \alpha_1 - \alpha_2 - \alpha_3 ) \right).
\end{align*}
where $ \Lambda $ is the Lobachevsky function
\begin{align*}
\Lambda(x) = - \int_0^x \log| 2 \sin(t) | \; dt.
\end{align*}
The angles $ \alpha_i $ and $ \beta_i $ are related to the triangle (in $ \widetilde{g} $ geometry) with edge lengths $ l_i $  and its 
circumcircle as shown in Figure (\ref{fig:tricirc}).

\begin{figure}[h]
\begin{tikzpicture}[scale = 2]
\coordinate (a) at (0.173648, 0.984808);
\coordinate (b) at (-0.939693, -0.34202);
\coordinate (c) at (0.642788, -0.766044);
\coordinate (c1) at (0.939693, -0.34202);
\coordinate (a1) at (0.642788, 0.766044);
\draw[scale=1,domain=0:360,smooth,variable=\x,black] plot ({cos(\x)},{sin(\x)});
\draw (a)--(b) node [above, left, midway] {\tiny{$ l_{3} $}};
\draw (b)--(c) node [below, midway] {\tiny{$ l_{2} $}};
\draw (c)--(a) node [right, midway] {\tiny{$ l_{1} $}};
\pic [draw, "\tiny{$\alpha_1$}",angle radius = .8cm] { angle = c--b--a};
\pic [draw, "\tiny{$\beta_1$}",angle radius = 1.03528cm] { angle = c1--c--a};
\pic [draw, "\tiny{$\beta_1$}",angle radius = 1.03528cm] { angle = c--a--a1};
\end{tikzpicture}\caption{}
\end{figure}\label{fig:tricirc}

Algebraically, the $ \alpha $ and $ \beta $ are related by
\begin{align*}
\alpha_i + \beta_j + \beta_k = \pi,
\end{align*}
and the $ \alpha $ can be computed explicitly from the edge lengths using the formulas
\begin{align*}
\tan \left( \frac{\alpha_i}{2} \right) = \begin{cases} \left( \frac{ (- l_{k} + l_{i} + l_{j}) (l_{k} + l_{i} - l_{j} ) }{( l_{k} - l_{i} + l_{j} ) (l_{k} + l_{i} + l_{j} )} \right)^{\frac{1}{2}} &\text{ if $ \widetilde{g} $ is euclidean}, \\
 \left( \frac{ \sinh(- l_{k} + l_{i} + l_{j}) \sinh(l_{k} + l_{i} - l_{j} ) }{\sinh( l_{k} - l_{i} + l_{j} ) \sinh(l_{k} + l_{i} + l_{j} )} \right)^{\frac{1}{2}}  &\text{ if $  \widetilde{g} $ is hyperbolic}, \\
\left( \frac{ \sin(- l_{k} + l_{i} + l_{j}) \sin(l_{k} + l_{i} - l_{j} ) }{\sin( l_{k} - l_{i} + l_{j} ) \sin(l_{k} + l_{i} + l_{j} )} \right)^{\frac{1}{2}} &\text{ if $ \widetilde{g} $ is spherical }.
\end{cases}
\end{align*}

\begin{theorem}[Variational Principle for Discrete Conformal Maps]
Every solution of the conformal mapping problem \ref{cmp} corresponds via (\ref{eq:confrmequiv}) to a critical point of the functional $ E^{\tilde{g}}_{\Theta} $. Conversely, a critical point of $ E^{\tilde{g}}_{\Theta} $ within the feasible region $ \mathcal{F} $ gives a solution to the conformal mapping problem \ref{cmp}.
\end{theorem}

Furthermore it was shown in \cite{SPS} for $ \tilde{g} $ euclidean, and in \cite{BPS} for $ \tilde{g} $ hyperbolic that the functional $ E^{\tilde{g}}_{\Theta} $ is convex in $ \mathcal{F} $, and can be extended to a convex function defined on $ \mathbb{R}^{|V|} $. This is crucial both theoretically and practically. The convexity implies that critical points of $ E^{\tilde{g}}_{\Theta} $ are global minima of the functional. Moreover, when $ \tilde{g} $ is hyperbolic, the functional is strictly convex and so has a unique minimum direction. When $ \tilde{g} $ is euclidean, the minimum of the functional is unique up to an overall scale.

Practically, for numerical calculations, this implies that one can utilize powerful numerical tools developed for convex optimization. 

In the case $ \tilde{g} $ is spherical, $ E^{\tilde{g}} $ is not convex, and solutions may not correspond to minima. In the context of the vortex equations, this corresponds to the fact that vortices with $ C > 0 $ do not correspond to minima of the Higgs action functional, and are not stable. Nonetheless, with some modification (see \cite{BSS}), the standard numerical methods can be succesfully employed.

On the other hand, the existence of solutions to the conformal mapping problem \ref{cmp} is more complicated. Solutions to the discrete conformal mapping problem may fail to exist due to violated triangle equalities (\ref{eq:tri}). On the other hand, if we regard the underlying surface $ S $ with constant curvature and conical singularities as the fundamental independent of its triangulation, then it is natural to consider re-triangulations.  Indeed, the theory of discrete conformal equivalence can be extended to combinatorially inequivalent triangulations (see \cite{BPS}). This generalization leads to progress with the existence problem. In particular, for compact surfaces, a discrete uniformization theorem has been proven in \cite{Luo1,Luo2}.

\section{Discrete Vortex Solutions}
In this section, we demonstrate how the discrete conformal mapping problem can be used to construct vortex solutions. Comparing the vortex equation in the form (\ref{eq:bapt}) to the conformal mapping problem \ref{ccmp}, it is clear how to proceed: given a Riemannian manifold, we choose a triangulation $ (\Sigma, l, g ) $ with $ g $ chosen so that $ C_g = C_0 $. The conical angles $ \Theta^i $ can then be calculated at each vertex. A solution to the vortex equation then corresponds to solving the conformal mapping problem \ref{cmp} with geometry $ \tilde{g} $ chosen so that $ C_{\tilde{g}} = C $, and identical conical angles $ \widetilde{\Theta}^i  = \Theta^i $ except at vortex centers, where  $ \widetilde{\Theta}^i = \Theta^i + 2 \pi n_i $.

\end{document}